\RequirePackage[2020-02-02]{latexrelease}
\documentclass[aps,prb,preprint,nopacs,superscriptaddress,]{revtex4}

\usepackage{graphicx}
\usepackage{verbatim}
\usepackage{mathrsfs}
\usepackage{gensymb}
\usepackage{color}
\usepackage{ulem}
\usepackage{times}

\usepackage{etoolbox}

\usepackage{amsmath,amsfonts,amssymb}
\usepackage{graphicx}

\begin{document}

\title{Nanorobotic actuator based on interlayer sliding ferroelectricity and field-tunable friction}
 
\author{Hechen Ren $^\dagger$\footnote[0]{$^\dagger$ Corresponding Author: ren@tju.edu.cn}}
\affiliation{Center for Joint Quantum Studies, Tianjin Key Laboratory of Low Dimensional Materials Physics and Preparing Technology, School of Science, Tianjin University, Tianjin 300350, China}
\affiliation{Joint School of National University of Singapore and Tianjin University, International Campus of Tianjin University, Binhai New City, 350207, Fuzhou, China}

\author{Jiaojiao Wang}
\affiliation{Center for Joint Quantum Studies, Tianjin Key Laboratory of Low Dimensional Materials Physics and Preparing Technology, School of Science, Tianjin University, Tianjin 300350, China}

\author{Wenxue He}
\affiliation{Center for Joint Quantum Studies, Tianjin Key Laboratory of Low Dimensional Materials Physics and Preparing Technology, School of Science, Tianjin University, Tianjin 300350, China} 

\begin{abstract}
Interlayer sliding ferroelectricity has been discovered in a variety of 2D materials with superb features such as atomic thickness, fast response, and fatigue resistance. So far, research on this phenomenon has been limited to fundamental physics and electronic applications, leaving its potential for electromechanical actuation unexplored. In this work, we design an atomic-scale actuator based on sliding ferroelectricity and field-tunable interfacial friction. With a prototype based on parallelly stacked bilayer h-BN sandwiched between gold contacts, we show how an alternating electric field can drive the bilayer into controlled crawling motions and how uniaxial strain can steer the crawl direction. Using numerical simulations, we demonstrate the actuator’s robust operation under a wide range of drive signals, friction scales, and frictional variations. We further provide experimental directions on how to realize field-tunable friction on h-BN interfaces. The wireless-ready actuation mechanism can be generalized to many 2D material systems possessing sliding ferroelectricity and integrated into flexible electronics platforms, opening new avenues in the development of intelligent nanorobotics.
\end{abstract}

\maketitle

\section*{Introduction}
\label{sec1}

Two-dimensional (2D) materials have emerged as a family of layered materials with rich physical properties and functionalities \cite{Novoselov2005a}, positioning them and their heterostructures as a competitive platform to build the next generation of integrated transistors \cite{Akinwande2014, Liu2024a}, memories \cite{Wang2022a, Chen2024a}, emitters \cite{Withers2015}, and sensors \cite{Oh2021}. As a late addition to their novel characteristics, interlayer sliding ferroelectricity occurs between weakly coupled layers and possesses atomic thickness, ultrafast response, and anti-fatiguing \cite{Bian2024}. Since its recent discovery in metallic $\text{WTe}_2$ \cite{Fei2018} and insulating hexagonal boron nitride (h-BN) \cite{Vizner_Stern_2021, Yasuda_2021, Woods2021}, this fascinating phenomenon has been discovered across a wide range of 2D materials such as $\text{MoS}_2$ \cite{Wang2022, Weston2022, Meng2022}, $\text{MoTe}_2$ \cite{Jindal2023}, $\text{In}_2\text{Se}_3$ \cite{Zhou2017}, and InSe \cite{Hu_2019, Sui2023a} among other materials and their heterostructures \cite{Zheng2020, Niu2022, Kang2023a}. Bringing a new wave of excitement in the interdisciplinary field of 2D materials known as slidetronics, sliding ferroelectricity has been zealously investigated for its domain-related physics \cite{McGilly2020} and for its applications in field-effect transistors \cite{Liu_2021} and non-volatile memory \cite{Gao2024}.

To date, an important aspect of 2D materials that has remained largely unexplored is its potent role in robotic actuation. Being piezoelectric, the switching of interfacial polarization is accompanied by the relative displacement between atomic layers. This presents an opportunity of using electric control of the polarization to dynamically drive the interlayer sliding. The actuation becomes drastically more useful when the layers can move continuously so that mesoscopic displacement with atomic precision can be achieved. Such a nanorobotic actuator is highly desirable in revolutionizing scientific apparatus \cite{Zhang2023a, Zhang2024}, medical instruments \cite{Yang2023b}, and everyday nanorobotic applications \cite{Wang2021d, Ye2023}. Recent efforts exploring the mobile aspect of van der Waals materials have used microscope tips to fold graphene \cite{Chen2019a} and pushed gold electrodes to slide across the surface of h-BN  \cite{Barabas2023}. These preliminary demonstrations reveal a promising use of this material family’s superlubricity in research settings, but the reliance on manual operations lacks the remote control that modern-day robotics are built upon. A fully electric-controlled, continuous-range crawler is what is needed to unleash layered materials’ true power in nanorobotics, but how to design such a crawler remains uncharted territory.

Current experimental setups are unfit for electromechanical actuation due to three systematic limitations. Firstly, most experiments study twisted or large samples where multiple domains coexist. This prevents the movement of an entire layer when the system is driven by a global field. Secondly, even for experiments containing single domains, the degeneracy rooted in the lattice symmetry offers neither a deterministic preference nor a statistical trend for the sliding to take a particular direction. As a result, repeated polarization switches induce either back-and-forth shifts or random walks in stacking configuration space instead of long-range, unidirectional actuation. Lastly, the device structure usually features uneven friction on the ferroelectric bilayer so that one layer is stuck to the surface of the substrate while the other layer predominantly slides. This prohibits the coordinated sliding motions both layers need to participate in for a nanobot to crawl.

In this work, we predict a new electromechanical effect where an alternating global electric field can drive continuous interlayer sliding and propose a nanorobotic crawler based on this mechanism. To shatter the above-mentioned limitations, our prototype design integrates elements of parallelly stacked bilayer h-BN, uniaxial strain, and field-tunable friction. By numerically simulating the sliding dynamics, we establish a suitable range of conditions where the crawler can properly function. We also provide concrete experimental directions on how to construct the needed field-tunable friction. Modeled after realistic parameters, our theory applies to a variety of materials where sliding ferroelectricity is present and opens new avenues for nanorobotics in 2D materials.

\section*{Results}
\label{sec2}
 
\subsection*{Nanorobotic actuator design based on interlayer sliding ferroelectricity}
\label{subsec2}

In bulk form, h-BN features AA’ stacking configuration which restores the centrosymmetry missing in its monolayer. The resulting electric polarization perpendicular to the sample plane averages to zero over the unit cell. When experimentalists stack one monolayer on top of another to make a parallel stack, this centrosymmetry is broken, and the two degenerate ground states feature AB and BA alignments. In AB stacking, the A atoms on the top layer (nitrogen) sit directly above the B atoms (boron) on the bottom layer, whereas the bottom layer’s A atoms (nitrogen) see no atoms directly above them (Fig. 1a). Due to charge redistribution, positive charge concentrates more toward these bottom nitrogen atoms, creating a net electric polarization over the unit cell pointing upward \cite{Vizner_Stern_2021}. We visualize the polarization in stacking configuration space in Fig. 1b \cite{Bennett2023}. Without loss of generality, let the origin $\textbf{\textit{s}} = (0, 0)$ represent an AB stacking and  $\textbf{\textit{s}} = (1, 0)$ represent the first BA stacking where the top layer slides one bond length to the right. The polarization is opposite for AB and BA stacking configurations and net zero for AA stacking. This allows us to select AB or BA stacking by controlling the interlayer polarization with an external perpendicular electric field. When the twist angle between the two layers remains zero and the whole bilayer consists of a single domain, the switching of electric polarization is naturally accompanied by a global slide between the layers by one bond length, as has been experimentally demonstrated \cite{Yasuda_2021}. 

One complication inherent in materials with hexagonal lattices is their rotational symmetry. For parallelly stacked h-BN, the switch between AB and BA stackings always presents three nearly degenerate directions to slide (Fig. 1a). To ensure directional actuation, we need to isolate one preferred sliding direction, which can be achieved by applying a uniaxial strain in-plane. This will break the $\textbf{\textit{C}}_3$ symmetry of the lattice and lower the energy along one of the AB-BA paths compared to the other two. Now, when an external field is applied in alternating sequences, the bilayer will slide along this preferred path and shift between two stacking configurations without wandering off. We will provide a more quantitative analysis of this effect in the next section. In real applications where the bilayer flake is kept small to maintain a single domain, sample geometry, and other environmental factors may also play the role of symmetry breaking or strain inducing.

Once we have unidirectional motion, we need to utilize the back-and-forth interlayer slide and coordinate which layer slides in a cyclic sequence so that crawling may happen. For example, when an AB stack slides into the new ground state of BA stacking, it can happen by any combination of the top layer sliding right and the bottom layer sliding left, and vice versa for a BA to switch to AB. An ideal design would have the top layer slide right during an AB-to-BA transition while the bottom layer slides right for a BA-to-AB transition (Fig. 1c). If we repeat this cycle by alternating the external field, and the layers move one at a time like two feet of a pedestrian, we will have accomplished the desired continuous-range motion of the bilayer relative to its surroundings.

In order for this coordinated footwork to happen, the frictional coefficients need to vary in phase with the polarization switching. For the top layer to slide right and the bottom layer to linger, the maximum static friction between the bottom layer and the bottom wall needs to surmount the electrostatic force driving the bottom layer to slide left. By Newton’s third law, the latter is equal to the force driving the top layer to slide right, which in turn must overcome the maximum static friction between the top layer and the top wall. For the next step, the order of the forces needs to be reversed for the bottom layer to move. This dictates that 1) the static frictions between each h-BN layer and their neighboring walls be comparable in scale to the polarization-switching force and 2) the frictions be tunable by the reversal of an external field. In our numeric simulations in the next section, we will see that condition 1 can be relaxed to an extent, especially given the polarization-switching force is proportional to the external field which can be tuned. As for condition 2, thanks to the studies accumulated on tribology, there are several candidate interfaces, which we shall detail in the discussion section. Here, we will use the example of the h-BN/Au interface to introduce our mechanism.

Figure 1c shows a crawling actuator design with two parallelly stacked h-BN layers sandwiched between gold electrodes. A major contributor to the frictional force between h-BN and Au comes from a polar charge separation with more positive charge concentrated in the h-BN layer and negative charge centered in the Au layer \cite{Baksi2019}. When the external field points up, friction between the top layer and the top wall decreases due to decreased charge separation at the Au/h-BN interface; on the contrary, friction between the bottom layer and the bottom wall increases. This asymmetry allows the top layer to slide against the bottom layer and the top wall. On the other hand, when the driving field points down, flipping the BA stack back to an AB stack, the top layer experiences more resistance than the bottom layer, so the bottom layer will slide right instead of the top layer sliding left. In this fashion, the two layers will alternate in sliding while maintaining their bulk overlap area, forming a true nanorobotic crawler. 

\par\quad\par
   
\begin{figure}[h]
\centering
\includegraphics[width=160mm]{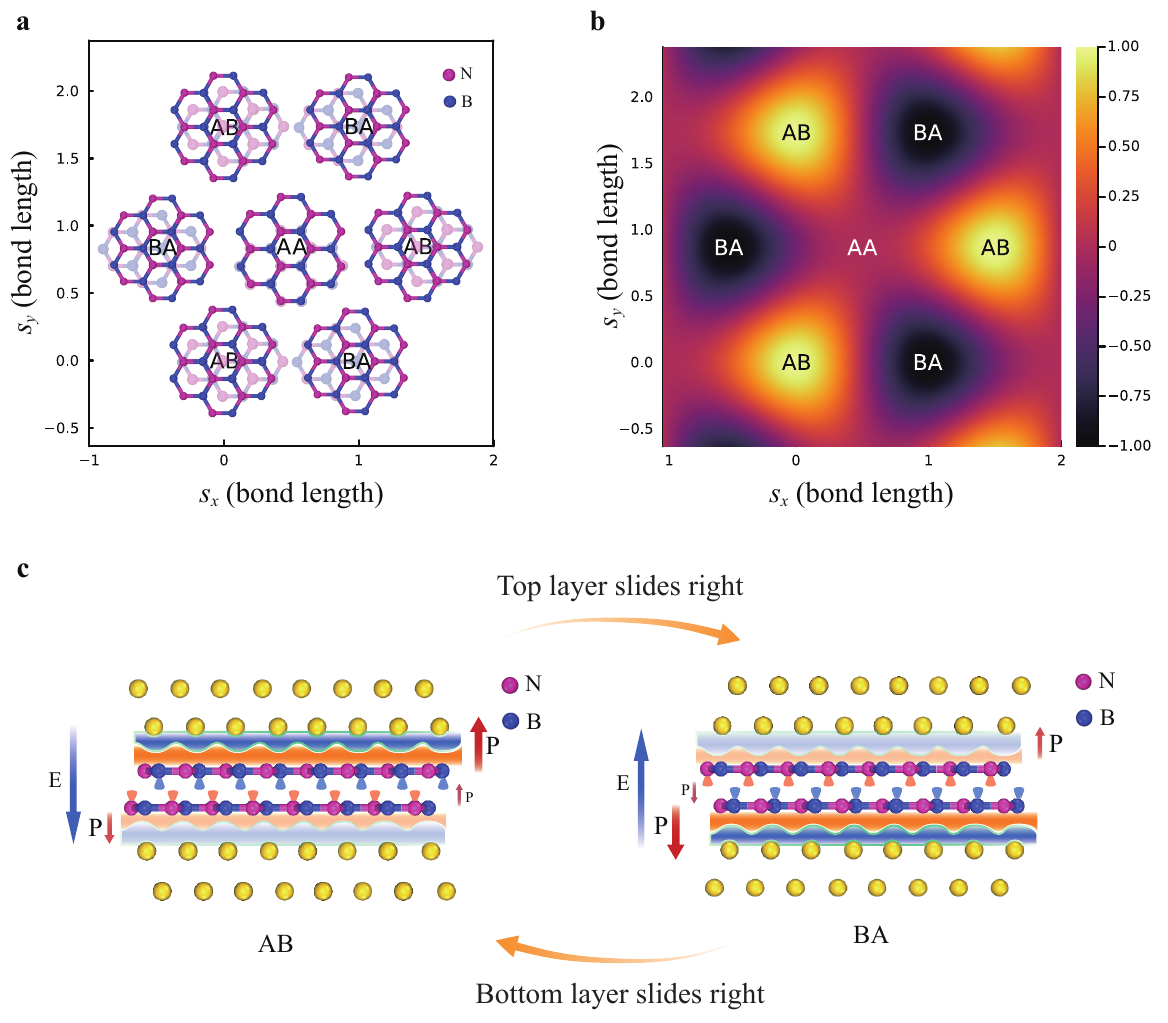}
\caption{
Schematic diagrams of the actuator's design. 
\textbf{a}	Stacking configurations of parallelly stacked h-BN in sliding space. The origin marks an AB stacking configuration, whereas the top layer sliding one bond length to the right (or in the 120-degree and 240-degree directions from the $x$-axis) gives a BA stacking configuration.
\textbf{b}	Normalized out-of-plane electric polarization in the same configuration space.
\textbf{c}	Crawling mechanism for the bilayer actuator driven by an alternating external field (blue arrows). 
}\label{fig1}
\end{figure}

\subsection*{Uniaxially strained bilayer h-BN in an electric field}
\label{subsec3}

To illustrate the symmetry-breaking effect and to estimate the polarization-switching force, we model uniaxially strained bilayer h-BN in an alternating electric field. Two components of electrostatic energy come into play in the sliding dynamics of the bilayer system. The first one is the interlayer interaction modeled by a Lennard-Jones (L-J) potential plus the Coulomb energy between atoms in different layers \cite{Vizner_Stern_2021}. 

\begin{equation}
E_0 = \frac{1}{2} \sum_{ij} \left[ 4 \epsilon \left(  \left( \frac{\sigma}{r_{ij}} \right) ^ {12} - \left( \frac{\sigma}{r_{ij}} \right) ^ {6} \right) + \frac{q_i q_j}{r_{ij}} \right]
\label{eq1}
\end{equation} 
where $\epsilon$ is the cohesive energy, $\sigma$ is the interlayer spacing scale, $q_{i,j} = \pm q$ are the partial charges of boron and nitrogen atoms, and $r_{ij}$ are the distances between atoms on different layers. Let $\textbf{\textit{s}}=(s_x, s_y)$ represent the top layer’s relative displacement from the bottom layer. Then we have $r_{ij} = \sqrt{ d^2 + (\Delta x_{ij} + s_x)^2 + (\Delta y_{ij} + s_y)^2}$ for each atomic pair between the top and the bottom layers, and the above interlayer energy $E_0$ can be expressed as a function of {\boldmath$s$}.

Figure 2a displays the combined L-J and Coulomb energy landscapes for a perfect honeycomb lattice. Using an interlayer spacing scale $\sigma = 3.3 \text{ \r{A}}$ and distance $d = 3.34 \text{ \r{A}} $, a cohesive energy ratio of $\epsilon / q^2 = 3 $ meV, and fit according to previous DFT calculations \cite{Constantinescu2013}, we obtain the stacking energy as a function of the relative sliding position. As shown in Fig. 2a, AA stacking features the highest energy surrounded by six energy minima representing AB and BA stackings. Line cuts in Fig. 2d show the potential profiles along the positive horizontal direction and the 60-degree direction counterclockwise from the $x$-axis. In pristine bilayer h-BN, both profiles feature an energy barrier above 8 meV around the AA valley and about 1 meV along a direct path connecting an AB and a BA valley.

The second part of the electrostatic energy comes from the net polarization $P$ in an external electric field $\textbf{\textit{E}}$, which depends both on the stacking configuration and the time-dependent external field. Now, a positive field favors BA stacking by lowering its electrostatic energy while raising that of AB stacking.

\begin{equation}
E_p = \textbf{\textit{P}} \cdot \textbf{\textit{E}}
\end{equation}

According to experimental results from single-domain samples, AB and BA configurations each possess a polarization of $2\times 10 ^{-12}$ C/m or 0.6 Debye/$\text{nm}^2$ and an electric field on the order of 0.1 V/nm suffices to flip the polarization \cite{Yasuda_2021}. Given the unit cell area of h-BN is 0.054 $\text{nm}^2$, this amounts to 0.032 Debye / B-N pair, which gives an energy difference of $\Delta E = 0.135 \text{ meV}$. This energy gain over a sliding distance of one bond length yields an average polarization-switching force on the scale of 0.15 pN. This puts an upper bound on the scale of the static force resisting the slide between AB and BA valleys, which comes from the combined L-J and Coulomb effects. It is an upper bound because the switching force also needs to overcome friction with surrounding layers. 

Given this small energy scale induced by a moderate electric field, the high barrier around AA stacking means polarization switching takes place by going from AB to BA directly without cycling through AA. Among the three choices of sliding directions, sliding backward into its original configuration is the most likely—the fact it has been selected the first time suggests it has the lowest energy among the initial three paths. Probabilistically, it will be energetically favorable among the three new paths from BA to AB as well. 

This accidental breaking of symmetry in practical systems may suffice to prefer the shifting between two fixed valleys in configuration space. However, to ensure the switching happens deterministically in a closed-loop fashion, we introduce a uniaxial tensile strain in the sample plane. Figure 2b shows the resultant $E_0$ under a 3\% strain of the hexagonal lattice in the $y$-direction, where the three-fold rotational symmetry $\textbf{\textit{C}}_3$ is broken. Linecuts show the sliding path connecting a pair of AB and BA configurations in the $x$-direction has a lower energy barrier than the other sliding directions (Fig. 2b, e). The energy barrier is also significantly lower than in free h-BN, putting the static resistance much closer to the experimental value. Increasing the amount of uniaxial strain to 5\% practically erases the energy barrier on the horizontal AB-BA path, as shown in Fig. 2c, f. Although neglecting secondary effects of the external field and strain on the polarization \footnote{To justify this omission, the former has been calculated to be on the order of under 5\% per 0.1 V/nm field \cite{Vizner_Stern_2021}.}, our model captures the effective role uniaxial strain can play in breaking the $\textbf{\textit{C}}_3$ symmetry and securing a preferred sliding path.

\par\quad\par
   
\begin{figure}[h]
\centering
\includegraphics[width=160mm]{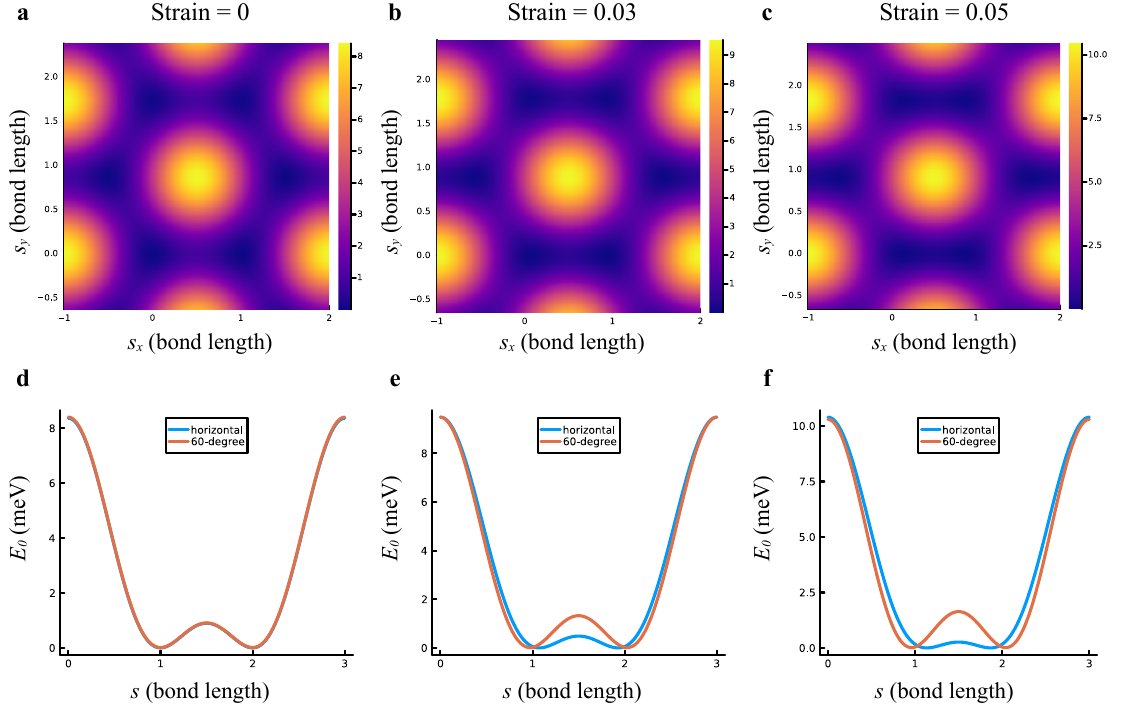}
\caption{
Uniaxial strain in bilayer h-BN breaks $\textbf{\textit{C}}_3$ symmetry and selects a sliding direction.
\textbf{a} Interlayer energy in stacking configuration space of pristine bilayer h-BN (in units of meV).
\textbf{b-c} Interlayer energies under 0.03 and 0.05 uniaxial tensile strains. The axes reflect the positions of the original lattice sites. 
\textbf{d-f} Linecuts of \textbf{a-c} in two directions, with horizontal line at $s_y=0$ and 60-degree line connecting $\textbf{\textit{s}}=(-1, 0)$ and $\textbf{\textit{s}}=(1/2, 3 \sqrt{3}/2)$. 
}\label{fig2}
\end{figure}

\subsection*{Sliding dynamics in alternating drive and tunable friction}
\label{subsec4}

Having eliminated the other two degenerate sliding directions via uniaxial strain, we can effectively treat the interplay between the static force, the polarization-switching force, and friction as a one-dimensional problem. Here, we again use the example of the h-BN/Au(111) interface. Previous DFT calculation suggests friction on the order of 4 pN per atomic pair with minimum external load. A recent experiment using atomic force microscopy measured this friction to be 800 nN/$\mu\text{m}^2$, which amounts to 0.04 pN per B-N pair. This discrepancy could be due to two reasons. Firstly, the DFT calculation uses highly lattice-matched Au(111) whereas the experiment includes deposited gold pads which may not be lattice-matched and could reduce friction. Secondly, the deposited gold surface isn’t necessarily atomically flat, which means the real contact area may be smaller than the nominal electrode size, further reducing friction. In summary, the range of friction we can achieve between h-BN and the gold surface is probably in a range between 0.04 pN and 4 pN per atomic pair depending on the material growth and interface treatment. If we open the search to other metals and layered materials for the walls, we have the liberty to choose from an even wider range of frictional forces as well as their change in an electric field.

Figure 3 illustrates how a crawler works with these elements. Using an average h-BN/Au friction of 1 pN and a static force profile calculated from the last section with 3\% strain (Fig. 2e), we can obtain the kinematic trajectories of both layers under an alternating electric field. Figure 3a gives an example at a 1-pN scale of polarization switching force, which requires 0.67 V/nm of external field and can easily be reached using direct gold contacts as walls. According to Newton’s third law, the interlayer forces—a combination of L-J potential, Coulomb energy, and polarization in the electric field—are equal in amplitude and opposite in direction on the two layers. These forces are plotted as blue lines in the bottom panels. For comparison, the maximum static friction on either layer is drawn in orange lines, which get modulated by the external field by a factor of 0.2. Whenever a layer isn’t in motion and its interlayer force magnitude is below its maximum static friction, it will remain still. However, when it is in motion, then the kinetic friction is taken to be 0.5 of the maximum static friction.

The top panels display the displacement of each layer as a function of time. During the first half of each drive cycle switching from an AB to a BA configuration, the top layer moves right while the bottom layer is held in place by its increased wall friction. In the second half, the top layer stays while the bottom layer moves right, switching back to the original AB configuration. During each cycle of alternating the external field in this fashion, each layer moves right by one bond length exactly, hence the whole actuator component has crawled within its sleeve by 1.44 $\text{ \r{A}}$. Figure 3a shows a 50-ps-period square-wave drive, while Fig. 3b shows a 100-ps-period drive with everything else being the same. After the initial 20-30 ps for each layer to settle, it doesn’t matter how long we wait before switching the field direction. This means we can make the crawler walk arbitrarily slowly—one bond length at a time for use cases like precise instrument control, or we can make it run ultrafast—up to a speed of 2.88 m/s with 20 GHz drive signals.

\par\quad\par   

\begin{figure}[h]
\centering
\includegraphics[width=160mm]{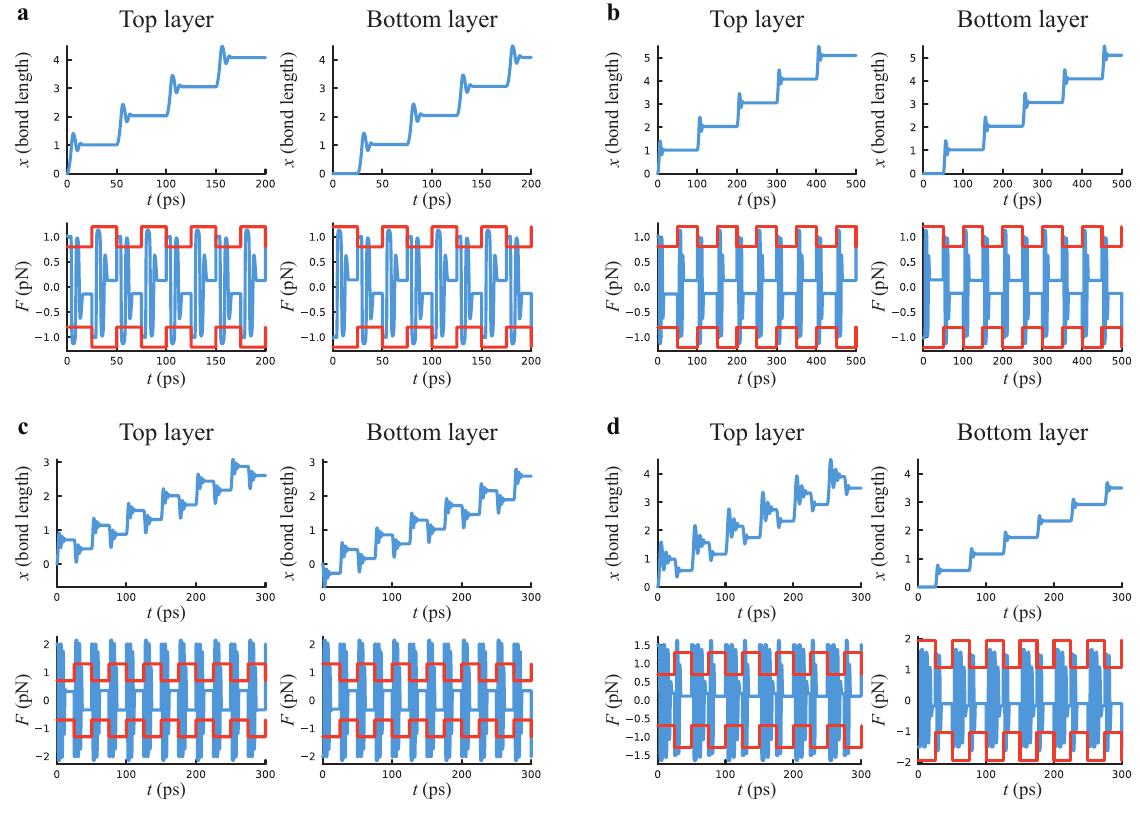}
\caption{
Operative modes of the bilayer sliding actuator under various conditions. 
\textbf{a}	A perfect functioning mode where the two layers alternate in sliding, using a drive period of 50 ps, a driving force scale of 1.0 pN, a friction scale of 1.0 pN, and a frictional variation of 0.2 on both sides. The top left and top right panels show the positions of the top and bottom layers respectively. The bottom left and bottom right panels show the combined driving force and the interlayer force on the top and bottom layers respectively. The maximum static frictions in both directions are drawn in orange lines for comparison. 
\textbf{b}	Another example of perfect functioning mode, using a drive period of 100 ps and all else the same as in \textbf{a}.
\textbf{c}	A slippery crawling mode when driving at a larger amplitude so that both layers simultaneously slide, using a driving force scale of 2.0 pN, a frictional variation of 0.3, and all else the same as in \textbf{a}.
\textbf{d} A functioning crawler with 50\% overall asymmetry between the two wall frictions, using a driving force scale of 1.5 pN, friction scales of 1.0 pN (top) and 1.5 pN (bottom), and all else the same as in \textbf{c}.
}\label{fig3}
\end{figure}

This simple design proves surprisingly robust against deviations from the ideal workflow. Figure 3c shows the actuator operating even when the switching force always surpasses the static friction. In this scenario, both layers slide during each cycle. Still, as long as field-driven asymmetry exists in the frictions on both sides, the sliding motions will be uneven between the layers, and the resulting motion is a slippery walk at a slower crawling speed. The even greater news is that the friction scale doesn’t have to exactly balance between the two walls either. Figure 3d shows an example where the overall friction on the top side is 50\% smaller than the bottom side. In this case, the top-layer h-BN moves more or even slips, but the actuator tolerates this imbalance and perseveres.

In Figures 4a-b, we map the crawling effect as a function of the polarization-switching force’s amplitude $F_\text{P}$ and the variation in friction $\Delta f$ tunable by the field. Fixing the driving period at 50 ps and the average friction scale at 1 pN (Fig. 4a) and 2 pN (Fig. 4b), we plot the bilayer’s final position at the end of 4 cycles (200 ps). We display the average of the top layer’s position and the bottom layer’s position, multiplied by a binary flag variable indicating if the two layers have separated by over one bond length relative to each other. This flag ensures that the crawling effect is zero when the intended actuation fails and the layers fall apart. In both two-dimensional plots, we see clear triangular boundaries within which crawling succeeds in the ideal sense, as represented by the case in Fig. 3a. The boundaries are set by the maximum static friction and the polarization switching force. Below the triangle, the switching force is too weak compared to the resistance on either side even when the field lowers the friction, so neither layer can slide. Above the triangle, we see faint regions where some crawling happens but slower than intended, which is the slippery scenario depicted in Fig. 3c.

In Fig. 4c, we fix the driving amplitude and vary the friction scale $f_s$ instead, allowing us to see the wide range of frictional parameters our design accepts. Again, a clear central region where the actuator crawls perfectly widens with more field tunability of the wall friction. In both types of maps (Fig. 4a-c), we see a bright boundary on one side of the operational region. We investigate this behavior in Fig. 4d, which suggests at this narrow resonance right before slipping begins to take place, the layers each leap more than one bond length per cycle, thus enhancing the crawling speed. Another thing to notice is the stripe features within the central zone of successful crawling. These slight deviations from perfect integer steps originate from the layers getting stuck near the lowest energy stacking ($s_x = 0  \text{ or } 1$) but never quite reaching there, as inhibited by the static friction. Within the central safe zone, the end position remains close enough to the integer steps to not jeopardize the actuator’s operation.

In Supplementary Fig. S1, we change the square-wave drive of the external electric field to a sine-wave drive and observe similar behaviors, suggesting the design’s general applicability to different drive types. We also map out the parameter space for the sine-wave drive (Supplementary Fig. S2). Interestingly, the narrow resonance peaks broaden into large regions comparable to the central safe zone. This presents an opportunity to trigger the actuator's leap mode in a more robust and accessible fashion.

\par\quad\par   
   
\begin{figure}[h]
\centering
\includegraphics[width=160mm]{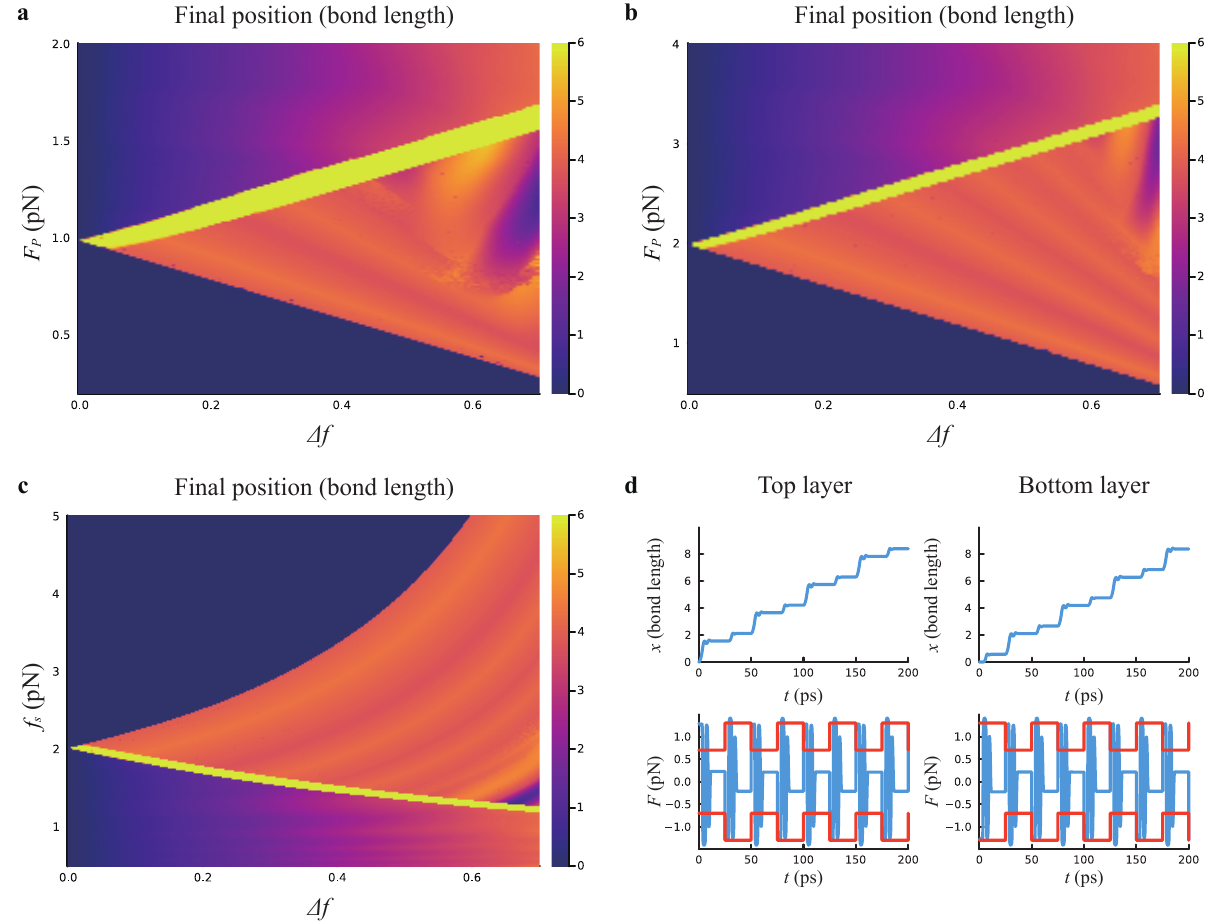}
\caption{
Actuation effect under different force scales and frictional variations. 
\textbf{a-b}  Crawling distance after 200 ps under a drive period of 50 ps as a function of driving amplitude and frictional variation (color scale is capped to preserve the main features). The friction scale is fixed at 1.0 pN in \textbf{a} and 2.0 pN in \textbf{b}.
\textbf{c}	Crawling distance after 200 ps under a drive period of 50 ps as a function of friction scale and frictional variation where the driving amplitude is fixed at 2.0 pN.
\textbf{d}	Resonant behavior where both layers slide two steps in each cycle. Panel layout follows the convention in Fig. 3.
}\label{fig4}
\end{figure}

\section*{Discussion}
\label{sec12}

Having illustrated the working principles of the nanorobotic crawler, let us now provide some realistic examples where field-tuned friction can be found at h-BN interfaces. The first example is the one we used above at h-BN/Au(111) interfaces. Here, because the friction is largely attributed to the vertical charge separation at the interface, and this dipole moment increases or decreases depending on the field direction, we can modulate the interfacial friction by reversing the field. The actual percentage of tunable friction at various h-BN/Au interfaces can be established both via first-principle calculations and via experimental investigations and is beyond the scope of this work. In searching for the optimal wall material, we can also replace Au with other materials which may provide larger, more tunable friction at their interfaces with h-BN. Another place where field-tunable friction has been demonstrated is graphene interfaces. A recent experiment offers evidence that friction at a graphene surface can be field-modulated between 0.01 nN and 0.1 nN over the area of an AFM tip \cite{Greenwood2023}. A separate simulation on the graphene/h-BN interface places their friction at up to 0.5 nN/$\text{nm}^2$, amounting to 27 pN/B-N pair \cite{Mandelli2017}. Both studies suggest graphene to be a promising wall material. A third avenue to engineer field-tunable friction is via Li-ion intercalation at h-BN interfaces, such as h-BN/graphene or h-BN/Au interfaces \cite{Pang2020}. Here, the mobile Li-ions serve as charge redistributors that contribute to the surface friction. Due to the separation of positive and negative charges surrounding them, these ions will reposition themselves in response to the direction of an external field, hence changing the interfacial friction. Finally, a fourth direction to explore is organic polymers or molecules whose friction can be field-tuned via relocation of ions or reorientation of polymers \cite{Gianetti2023}. All these approaches and combinations among them are worth investigating as field-tunable friction and electroadhesion are mighty goals in their own right for nanoscale research and nanorobotic designs \cite{Guo2020}.

Next, we would like to comment on the steering of the actuator. In this current design, the nanobot moves in the direction aligned with the sliding, which is determined by the lattice orientation and the uniaxial strain. In order to have a steering wheel on the nanobot—to direct its movement in every step as a programmable variable—we need to be able to change the strain direction in real time. Since the strain merely serves to break the degeneracy and doesn’t need to be very large, this could be achieved in multiple ways. One traditional design is to have piezoelectric contacts along the periphery of the bilayer. Groups of contacts on opposite sides can then be actuated electrically to stretch or relax the bilayer in designated ways. Another way is to integrate this actuator on a flexible electronic platform, where the curving of the surface around one direction could lead to a tensile strain perpendicular to it. This would then favor crawling along the curving axis. More creative techniques such as those exploiting surface acoustic waves \cite{Nie2023} or intercorrelated in-plane and out-of-plane ferroelectricity \cite{Hu_2019, Liang2021a, Bennett2023} are reserved for future works to explore.

We would also like to emphasize the completeness of the crawling motion to serve as the basis for planar navigation. The group generated by linear combinations of crawling in the three AB-BA directions spans the full $\mathbb{R}^2$ space. This means with sequential turning, the nanobot can crawl anywhere in the two-dimensional plane. A pictorial proof is provided in Supplementary Fig. S3, where the entire $\mathbb{R}^2$ space is divided by the crawling axes into three 120-degree sectors. Any point on the axes can be reached via a single crawl. Any other planar displacement can be accomplished by two crawling steps along the axes bordering its sector. The reason we have three directions instead of two to cover the $\mathbb{R}^2$ plane is so that all the coefficients stay non-negative to respect the actuator’s chiral nature.

Finally, we would like to point out the semi-quantitative nature of our numeric simulations, which serve the purpose of conceptual illustration. For example, the tunability of interfacial friction may also vary with the driving amplitude. However, since the exact tunability varies in different material systems, we study the actuation’s dependence on the tunability as a free parameter and do not separately model how much it depends on the electric field. More precise parameters could be extracted from first-principle calculations and experimental measurements. The bottom line is that the relevant parameters such as stacking energy, polarization, and friction all remain universal given the material interface. Therefore, standard operating protocols for actuation in any direction over any distance can be calibrated once and for all. 

In conclusion, we have designed a new atomic-scale actuator based on sliding ferroelectricity and field-tunable interfacial friction. Under an alternating electric field, the strained bilayer can be driven into a cyclic sliding sequence that results in the crawling motion of the world’s thinnest nanobot. Modeling after real material parameters of h-BN, we simulate the kinetic motion of the actuator and establish a wide range of drive conditions, friction scales, and frictional variations under which the actuator operates. It is worth noting that parallelly stacked h-BN is by no means the only candidate; the slidetronics-based actuation mechanism we present here can be generalized to a variety of 2D material systems we have hitherto discovered or have yet to discover. The global field control without requirements on the frequency or phase opens up possibilities of untethered soft robots \cite{Jung2024}. Going forward, wireless signals can be received and amplified to actuate the nanobot, making it easy for its future development to adopt data-driven modeling and control \cite{Yasa2023}. Furthermore, it can be readily integrated into flexible-electronics platforms with sensors, memory, and processors \cite{Akinwande2014, Liu2024a}. With the addition of an atomic-precision, steerable actuator, 2D-material-based nanorobotics could harness the full power of reinforcement learning and mark an exciting new era for intelligent machinery on the atomic scale.

\section*{Methods}
\label{sec11}

Numeric calculations were scripted and visualized in the programming language Julia. For the two-dimensional plot of electric polarization, we adopted the formula in \cite{Bennett2023} with values taken from experiments on single-domain samples \cite{Yasuda_2021}. For the illustration of L-J potential and Coulomb energy in stacking configuration space, we used a small top layer with 9 B-N pairs sliding across a nearly infinite bottom layer with 10201 B-N pairs to avoid finite-size effects, and we scaled our unstrained energy profile by a factor of 0.75 to match the values calculated in DFT \cite{Constantinescu2013}. We differentiated the resulting 0.03-strain energy along the horizontal sliding direction to generate the static force profile for subsequent simulations, and we used a Fermi-Dirac distribution function to model nearly constant driving forces which quickly approaches zero near the equilibrium point (AB or BA stacking depending on the external field). The exact function of the driving force is subject to the material system’s dynamic charge redistribution in an external field during a sliding process, which is beyond the accuracy captured by this conceptual model. In all kinematic calculations, the time step was set at 0.001 ps, and the mass of the B-N pair was taken to be $4.1 \times 10^{-26}$ kg.

\clearpage

\setcounter{figure}{0}
\renewcommand{\figurename}{Fig.}
\renewcommand{\thefigure}{S\arabic{figure}}

\begin{figure}[h]
\centering
\includegraphics[width=165mm]{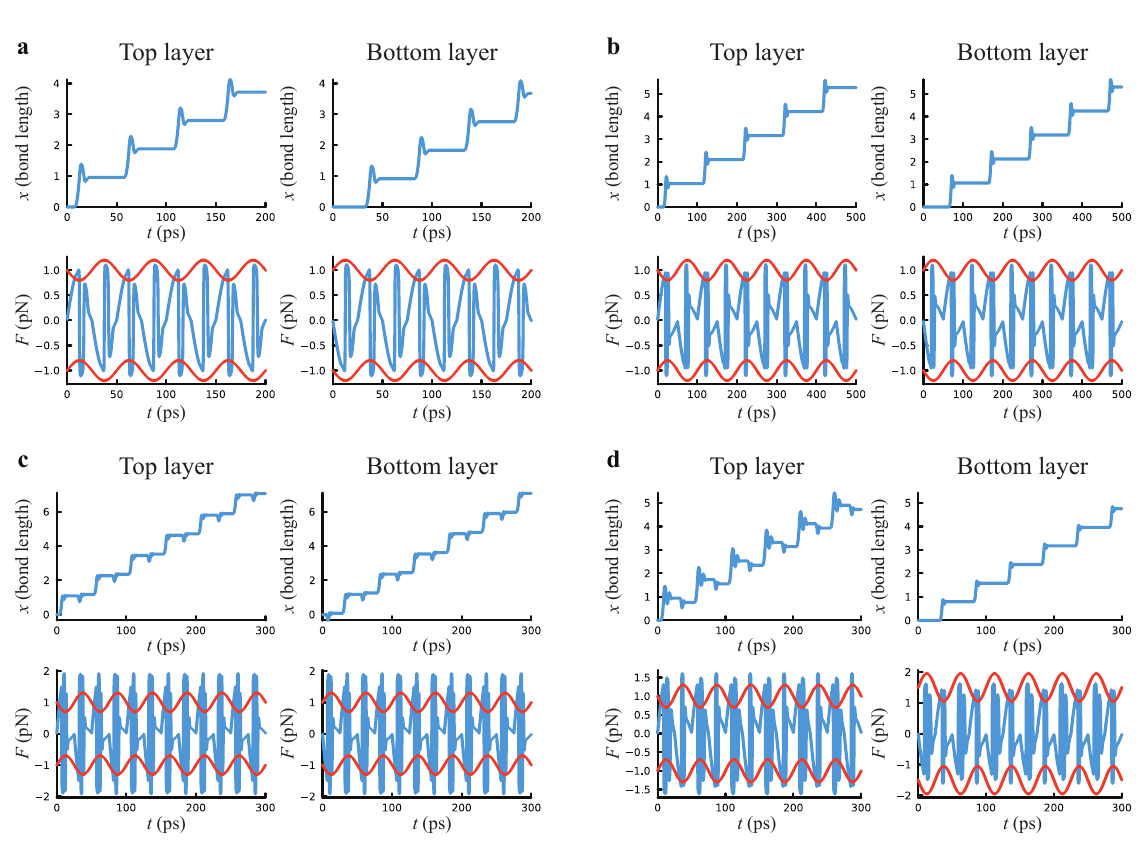}
\caption{
Operative modes of the bilayer sliding actuator using sine-wave drive signals. 
\textbf{a}	A perfect functioning mode where the two layers alternate in sliding, using a drive period of 50 ps, a driving force scale of 1.0 pN, a friction scale of 1.0 pN and frictional variation of 0.2 on both sides. Panel layout follows the convention in Fig. 3 of the main text.
\textbf{b}	Another example of the perfect functioning mode, using a drive period of 100 ps and all else the same as in \textbf{a}.
\textbf{c}	A resonant mode using a driving force scale of 2.0 pN, a frictional variation of 0.3 and all else the same as in \textbf{a}. Notice this same combination produces a slippery actuation with square-wave drive (Fig. 3c).
\textbf{d} A functioning crawler with 50\% overall asymmetry between the two wall frictions, using a driving force scale of 1.5 pN, friction scales of 1.0 pN (top) and 1.5 pN (bottom), and all else the same as in \textbf{c}.
}\label{figS1}
\end{figure}

\begin{figure}[h]
\centering
\includegraphics[width=165mm]{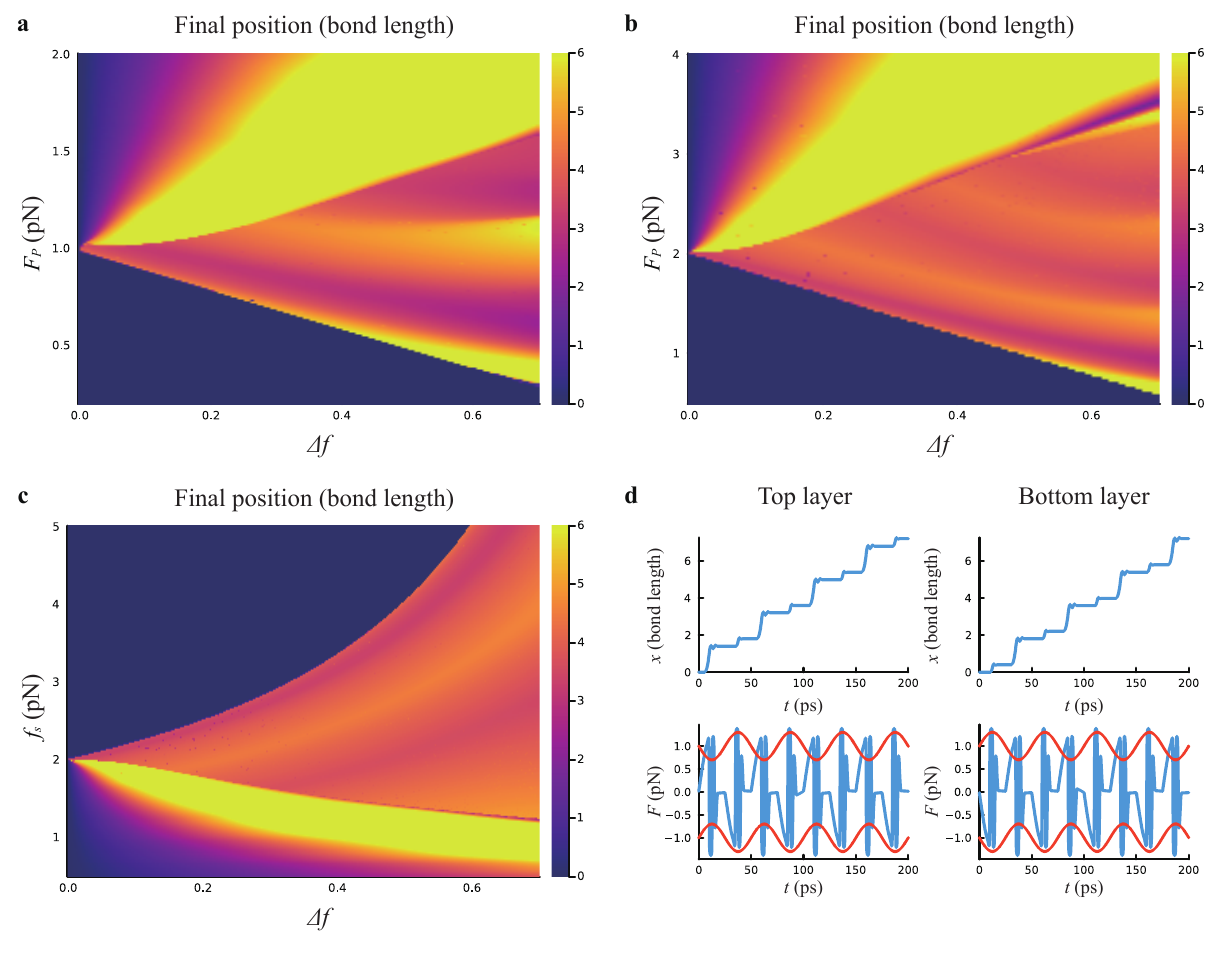}
\caption{
Actuation effect under different force scales and frictional variations. 
\textbf{a-b} Crawling distance after 200 ps under a drive period of 50 ps as a function of driving amplitude and frictional variation (color scale is capped to preserve the main features). The friction scale is fixed at 1.0 pN in \textbf{a} and 2.0 pN in \textbf{b}.
\textbf{c}	Crawling distance after 200 ps under a drive period of 50 ps as a function of friction scale and frictional variation where the driving amplitude is fixed at 2.0 pN.
\textbf{d}	Resonant behavior where both layers slide more than one bond length in each cycle. Panel layout follows the convention in Fig. 3.
}\label{figS2}
\end{figure}

\begin{figure}[h]
\centering
\includegraphics[width=125mm]{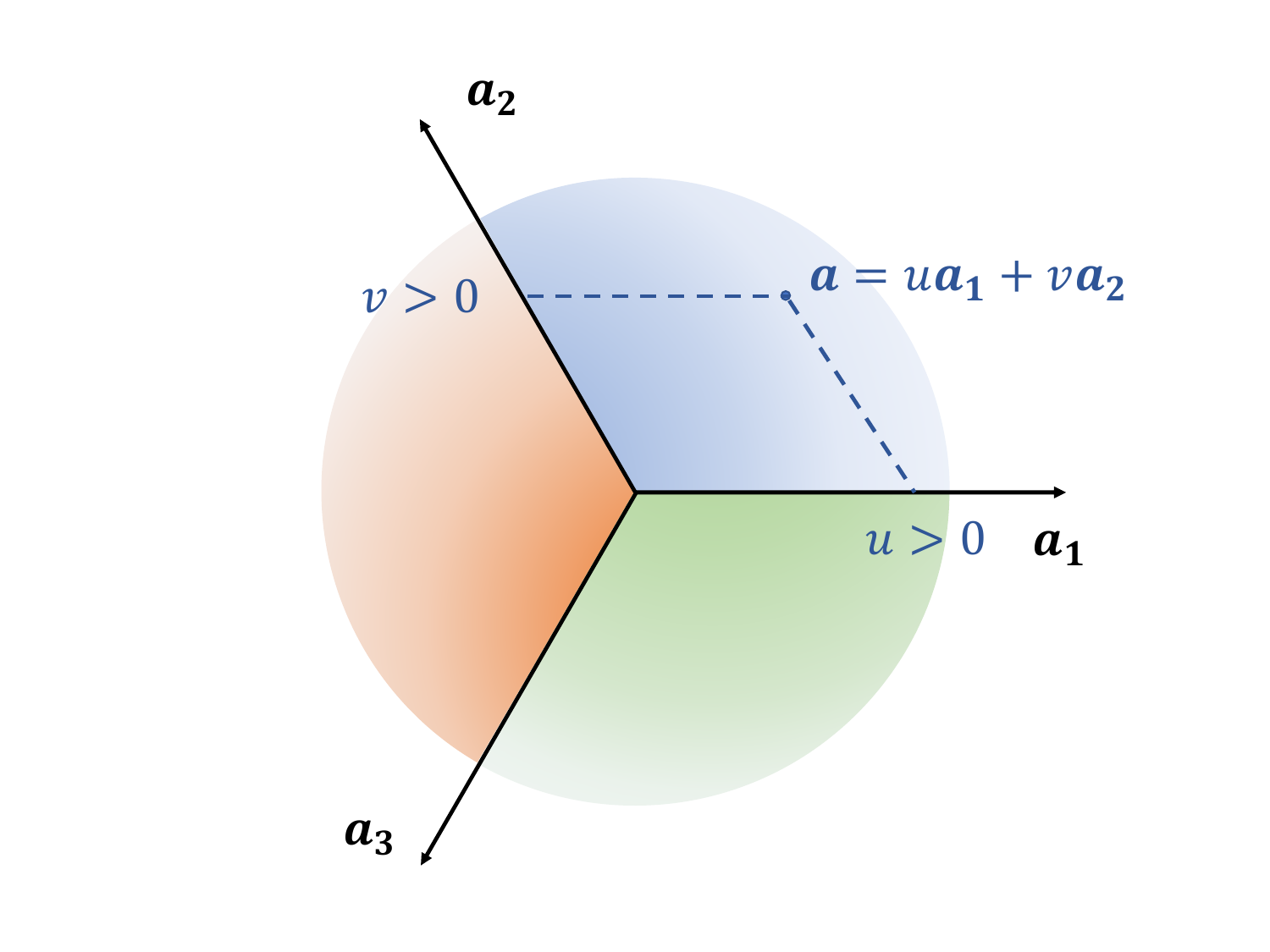}
\caption{
Completeness of the three basic crawling motions to reach all points in $\mathbb{R}^2$ space. Any point on the axes can be reached via a single crawl. Any point in the upper right sector (shaded in blue) can be decomposed into a linear combination of a crawl in the $\textbf{\textit{a}}_1$ direction and one in the $\textbf{\textit{a}}_2$ direction; any point in the lower right sector (green) can be decomposed into a linear combination of a crawl in the $\textbf{\textit{a}}_1$ direction and one in the $\textbf{\textit{a}}_3$ direction; any point in the left sector (orange) can be decomposed into a linear combination of a crawl in the $\textbf{\textit{a}}_2$ direction and one in the $\textbf{\textit{a}}_3$ direction. 
}\label{figS3}
\end{figure}

\end{document}